\documentclass[showpacs,showkeys,byrevtex,twocolumn,pra]{revtex4}%
\pdfoutput=1
\usepackage{amsfonts}
\usepackage{amsmath}
\usepackage{amssymb}
\usepackage{graphicx}
\usepackage{verbatim}%
\providecommand{\U}[1]{\protect\rule{.1in}{.1in}}

\begin{document}
\preprint{ }
\title[ ]{Protecting Quantum Information with Entanglement and Noisy Optical Modes}
\author{Mark M. Wilde}
\email{mark.wilde@usc.edu}
\author{Todd A. Brun}
\affiliation{Center for Quantum Information Science and Technology, Communication Sciences
Institute, Department of Electrical Engineering, University of Southern
California, Los Angeles, California 90089, USA}
\keywords{operator quantum error correction, stabilizer formalism, entanglement-assisted
quantum error correction, continuous variables, linear-optical quantum computation}
\pacs{03.67.-a, 03.67.Hk, 42.50.Dv}

\begin{abstract}
We incorporate active and passive quantum error-correcting techniques to
protect a set of optical information modes of a continuous-variable quantum
information system. Our method uses ancilla modes, entangled modes, and gauge
modes (modes in a mixed state) to help correct errors on a set of information
modes. A linear-optical encoding circuit consisting of offline squeezers,
passive optical devices, feedforward control, conditional modulation, and
homodyne measurements performs the encoding. The result is that we extend the
entanglement-assisted operator stabilizer formalism for discrete variables to
continuous-variable quantum information processing.

\end{abstract}
\volumeyear{2007}
\volumenumber{ }
\issuenumber{ }
\eid{ }
\date{\today}
\received{\today}

\revised{}

\accepted{}

\published{}

\startpage{1}
\endpage{ }
\maketitle

\section{Introduction}

Quantum computers and quantum communication systems will employ a variety of
techniques to protect quantum information from the negative effects of
decoherence \cite{PhysRevA.52.R2493,PhysRevLett.77.793,PhysRevA.54.1098,
thesis97gottesman,PhysRevLett.78.405,ieee1998calderbank}. \textit{Active}
quantum error-correcting techniques use multi-qubit measurements to learn
about quantum errors and correct for these errors
\cite{PhysRevLett.79.3306,mpl1997zanardi,PhysRevLett.81.2594}.
\textit{Passive} techniques exploit the symmetry of noisy quantum processes so
that the quantum information we wish to protect remains invariant under the
action of the noise
\cite{kribs:180501,qic2006kribs,poulin:230504,isit2007brun,hsieh:062313}.

\begin{figure}
[ptb]
\begin{center}
\includegraphics[
natheight=4.766800in,
natwidth=10.253200in,
height=1.5324in,
width=3.2846in
]%
{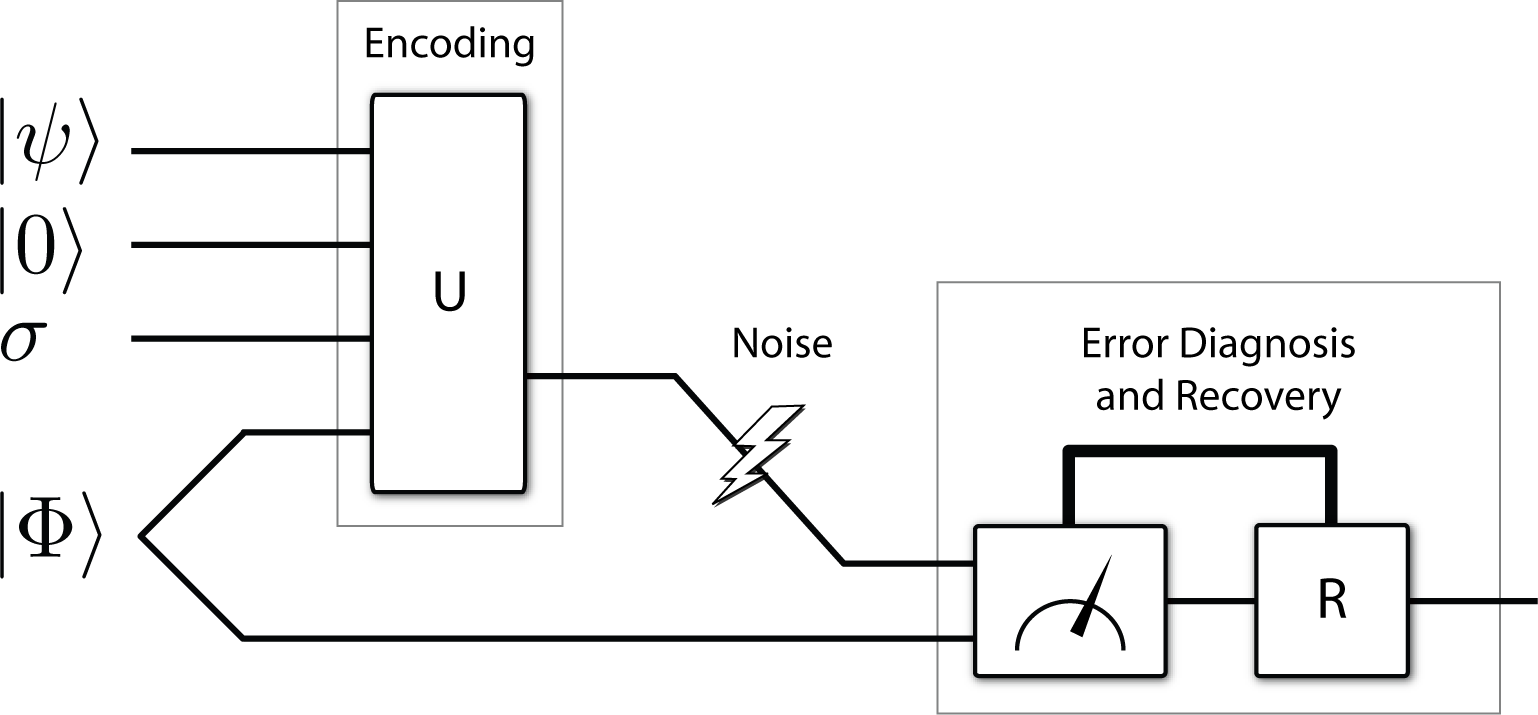}%
\caption{The operation of an $\left[  \left[  n,k;r,c\right]  \right]  $
entanglement-assisted operator quantum error-correcting code. The sender
begins with a set of $k$ information qubits in state $\left\vert
\psi\right\rangle $, $n-k-r-c$ ancilla qubits in state $\left\vert
0\right\rangle $, $r$ gauge qubits in a mixed state $\sigma$, and $c$ ebits in
state $\left\vert \Phi\right\rangle $. She encodes her information qubits with
the help of the ancilla qubits, gauge qubits, and her half of the ebits. The
receiver performs $n-k-r$\ measurements on all of the qubits to diagnose
errors. The $n-k-r$ measurements correspond to the\ $n-k-r$ ancilla qubits and
ebits. The code also has passive error-correcting capability due to the use of
ancilla qubits and gauge qubits.}%
\label{fig:eao-code}%
\end{center}
\end{figure}

We can classify the techniques according to the resources they employ for
quantum redundancy: ancilla qubits, entangled qubits (ebits), or gauge qubits
(qubits allowed to be noisy). A general technique for additive quantum error
correction is the entanglement-assisted operator stabilizer formalism
\cite{isit2007brun,hsieh:062313}---it employs ancilla qubits, ebits, and gauge
qubits for quantum redundancy. Figure~\ref{fig:eao-code} highlights the
operation of an entanglement-assisted operator code.%

Continuous-variable quantum information is an alternative to discrete-variable
quantum information and has become increasingly popular for quantum computing
and quantum communication \cite{book2003braunstein,revmod2005braunstein}.
Experimentalists have performed many \textquotedblleft
proof-of-concept\textquotedblright\ experiments
\cite{science1998furusawa,mizuno:012304,PhysRevLett.88.047904} that implement
most of the basic protocols in continuous-variable quantum information theory
\cite{PhysRevA.49.1473,prl1998braunstein,optb1999ban,pra2000braunstein}.
Continuous-variable experiments are less difficult to perform than
discrete-variable ones because they do not require single-photon sources and
detectors and usually require linear optical devices only---offline squeezers,
passive optical devices, feedforward control, conditional modulation, and
homodyne measurements. An offline squeezer is a device that prepares a
standard squeezed state for use in an optical circuit and an online squeezer
is a nonlinear optical device used in an optical circuit. It is possible to
simulate an online squeezer using a linear-optical circuit \cite{filip:042308}%
. The recent proposal \cite{filip:042308}\ and experimental implementation of
an online squeezer \cite{arx2007furusawa} and a quantum nondemolition
interaction \cite{conf2007furusawa}\ using linear optics should further
increase the popularity of continuous-variable quantum information processing.
Several continuous-variable quantum protocols use this scheme
\cite{wilde:060303,arx2007wildeCC,pra2007wildeEA}\ and many more protocols
should benefit from this technique.

Error correction is necessary for a continuous-variable quantum device to
operate properly. Several authors have suggested methods for error correction
of continuous-variable quantum information
\cite{prl1998braunstein_error,PhysRevLett.80.4088,nat1998braun,pra2007wildeEA,arx2007niset}%
. Some of these schemes
\cite{prl1998braunstein_error,PhysRevLett.80.4088,nat1998braun,pra2007wildeEA}%
\ are vulnerable to small displacement errors that occur in a
continuous-variable quantum system \cite{PhysRevA.64.012310}. They operate
well only when the squeezing of optical parametric oscillators is high and the
homodyne detectors are high efficiency. Experimentalists thus have a difficult
technological challenge to overcome if continuous-variable quantum devices are
to be practical. Nevertheless, these continuous-variable error correction
schemes should prove useful as a testbed for theoretical ideas even if the
final form of a quantum computer is not a continuous-variable optical device.

In this paper, we develop the entanglement-assisted operator stabilizer
formalism for continuous-variable quantum systems. Our continuous-variable
error correction scheme incorporates several forms of quantum redundancy:
ancilla modes, entangled modes, and gauge modes. The benefit of our theory is
that we can incorporate passive error-correction capability with a subsystem
structure while still have the benefits of an entanglement-assisted code.
Incorporating this subsystem structure may help in passively mitigating the
effects of the small displacement errors that plague continuous-variable
quantum systems.

We first briefly review the known techniques for discrete-variable quantum
error correction. The next section discusses a canonical entanglement-assisted
operator continuous-variable code and illustrate it with a quantum parity
check matrix. We show how a local unitary relates an arbitrary
entanglement-assisted operator code to the canonical one. Finally, we remark
how a linear-optical circuit can perform the encoding operations using the
techniques in Ref. \cite{braunstein:055801}\ or Ref. \cite{pra2007wildeEA}.

\section{Discrete-Variable Quantum Error Correction Techniques}

We first review the different techniques for protecting discrete-variable
quantum information. Each of the techniques falls into a class based on the
resources it employs for quantum redundancy. Ancilla qubits provide both
active and passive error-correcting capability, ebits provide active
error-correcting capability, and gauge qubits provide passive error-correcting
capability. A code is a subspace code if it uses only ancilla qubits or ebits,
and it is a subsystem or operator code \cite{PhysRevLett.84.2525}\ if it uses
gauge qubits in addition to ancilla qubits or ebits. Stabilizer codes employ
ancilla qubits only, operator codes employ ancilla qubits and gauge qubits,
entanglement-assisted codes employ ancilla qubits and ebits, and
entanglement-assisted operator codes employ ancilla qubits, ebits, and gauge
qubits. We review each of the above methods in more detail below.

An $\left[  \left[  n,k\right]  \right]  $ stabilizer code corrects errors
actively and passively by encoding $k$ information qubits with the help of
$n-k$\ ancilla qubits \cite{thesis97gottesman}. The formalism operates in the
Heisenberg picture by tracking a set of $n-k$ operators that stabilize the $k$
encoded information qubits. These $n-k$ operators equivalently correspond to
logical Pauli $Z$ operators for the $n-k$ ancilla qubits. The receiver
measures the $n-k$ operators corresponding to the encoded ancilla qubits to
diagnose errors. These $n-k$ measurements learn only about quantum errors that
occur and do not learn anything about the state of the $k$ information qubits.
Stabilizer codes are active because they employ measurements to learn about
the error and correct the encoded information qubits based on the result of
the measurements. They are also passive because the $n-k$ operators
corresponding to the ancilla qubits form a basis for errors that the code
corrects passively.

Operator quantum error correction unifies active/passive stabilizer subspace
coding techniques and passive subsystem techniques by respectively employing
ancilla qubits and gauge qubits for quantum redundancy
\cite{kribs:180501,qic2006kribs,kribs:042329,nielsen:064304,poulin:230504,isit2008aly}%
. An $\left[  \left[  n,k;r\right]  \right]  $ operator code encodes a set of
$k$ information qubits with the help of $n-k-r$ ancilla qubits and $r$ gauge
qubits. It operates similarly to a stabilizer code because the receiver
measures the logical Pauli $\bar{Z}$ operators corresponding to the $n-k-r$
ancilla qubits to diagnose and correct some of the errors. The logical Pauli
$\bar{X}$ and $\bar{Z}$ operators corresponding to the $r$ encoded gauge
qubits form a basis for the errors that the code passively corrects. Operator
codes have advantages over stabilizer codes to the degree that they might
allow us reduce the number of measurements and corrections we have to perform
to diagnose the error.

Entanglement-assisted codes correct errors both actively and passively by
employing ancilla qubits and ebits for quantum redundancy
\cite{arx2006brun,science2006brun,isit2007brun,hsieh:062313,pra2007wildeEA}.
The technique assumes that a sender and receiver share pure noiseless
entanglement (a set of ebits) prior to communication. An $\left[  \left[
n,k;c\right]  \right]  $ entanglement-assisted code encodes $k$ information
qubits with the help of $n-k-c$ ancilla qubits and $c$ ebits. Methods exists
to determine the optimal number of ebits that a given code requires
\cite{arx2008wildeOEA}. The crucial assumption in the entanglement-assisted
paradigm is that noise does not act on the receiver's half of the ebits. The
receiver measures the logical Pauli $\bar{Z}$ operators corresponding to the
$n-k-c$ encoded ancilla qubits and the logical Pauli $\bar{X}^{A}X^{B}$ and
$\bar{Z}^{A}Z^{B}$ operators corresponding to the $c$ encoded ebits to
diagnose errors. Operators with an $A$ superscript correspond to Paulis acting
on the sender's side and those with a $B$ superscript correspond to a Pauli
acting on the receiver's side. A benefit of entanglement-assisted coding is
that we can import any classical block or convolutional code for use in
quantum error correction \cite{science2006brun,arx2007wildeCED,prep2007wilde}.
Additionally, a source of pre-established entanglement boosts the rate of an
entanglement-assisted code. We can produce an $\left[  \left[  n,k;c\right]
\right]  $ entanglement-assisted stabilizer code from an $\left[  \left[
n,k\right]  \right]  $\ stabilizer code by replacing $c$ of the unencoded
ancillas of the stabilizer code with $c$ ebits. The resulting
entanglement-assisted code is more powerful than the original stabilizer code
because it corrects a larger set of errors.

Entanglement-assisted operator quantum error correction combines the benefits
of all of the above techniques by employing ancilla qubits, ebits, and gauge
qubits for quantum redundancy \cite{isit2007brun,hsieh:062313}. An $\left[
\left[  n,k;c,r\right]  \right]  $ entanglement-assisted operator quantum
error-correcting code encodes a set of $k$\ information qubits with the help
of $n-k-r-c$ ancilla qubits, $r$ gauge qubits, and $c$ ebits. This technique
is the one of the most powerful known techniques for quantum error correction
because it employs a large variety of resources for encoding.

\section{Continuous-Variable Mathematical Preliminaries}

We review a few mathematical preliminaries that are necessary for
continuous-variable quantum information processing before proceeding with our
theory of error correction (See Ref.~\cite{pra2007wildeEA}\ for a more
detailed review.)

The displacement operators are the most important for continuous-variable
quantum information processing. Let $X\left(  x\right)  $\ denote a
single-mode position-quadrature displacement by $x$ and let $Z\left(
p\right)  $ denote a single-mode momentum-quadrature \textquotedblleft
kick\textquotedblright\ by $p$ where%
\begin{align}
X\left(  x\right)   &  \equiv\exp\left\{  -i\pi x\hat{p}\right\}  ,\nonumber\\
Z\left(  p\right)   &  \equiv\exp\left\{  i\pi p\ \hat{x}\right\}  .
\end{align}
Operators $\hat{x}$ and $\hat{p}$ are the position-quadrature and
momentum-quadrature operators respectively with canonical commutation
relations $\left[  \hat{x},\hat{p}\right]  =i$. We can extend the above
definitions to multimode displacement operators with a map $\mathbf{D}$ as
follows%
\begin{equation}
\mathbf{D}\left(  \mathbf{u}\right)  \equiv\exp\left\{  i\sqrt{\pi}%
{\textstyle\sum\limits_{i=1}^{n}}
\left(  p_{i}\hat{x}_{i}-x_{i}\hat{p}_{i}\right)  \right\}  ,
\label{eq:map_symp_hw}%
\end{equation}
where $\mathbf{u}=\left(  p_{1},\ldots,p_{n},x_{1},\ldots,x_{n}\right)
\in\mathbb{R}^{2n}$ and the set of canonical operators $\hat{x}_{i},\hat
{p}_{i}$ for all $i\in\left\{  1,\ldots,n\right\}  $ have the canonical
commutation relations (in units where $\hbar=1$):%
\begin{align*}
\left[  \hat{x}_{i},\hat{x}_{j}\right]   &  =0,\\
\left[  \hat{p}_{i},\hat{p}_{j}\right]   &  =0,\\
\left[  \hat{x}_{i},\hat{p}_{j}\right]   &  =i\delta_{ij}.
\end{align*}
We also write $\mathbf{u}=\left(  p_{1},\ldots,p_{n}\ |\ x_{1},\ldots
,x_{n}\right)  $ where the vertical bar separates the momentum-quadrature
\textquotedblleft kick\textquotedblright\ parameters from the
position-quadrature displacement parameters. Let%
\begin{align}
\mathbf{X}\left(  \mathbf{x}\right)   &  \equiv X\left(  x_{1}\right)
\otimes\cdots\otimes X\left(  x_{n}\right)  ,\nonumber\\
\mathbf{Z}\left(  \mathbf{p}\right)   &  \equiv Z\left(  p_{1}\right)
\otimes\cdots\otimes Z\left(  p_{n}\right)  ,
\end{align}
so that $\mathbf{D}\left(  \mathbf{u}\right)  $ and $\mathbf{Z}\left(
\mathbf{p}\right)  \mathbf{X}\left(  \mathbf{x}\right)  $ are equivalent up to
a global phase. Another map $\mathbf{M}$ proves to be useful in our theory,
where%
\begin{equation}
\mathbf{M}\left(  \mathbf{u}\right)  \equiv\mathbf{u\cdot\hat{R}}^{n},
\label{eq:map_M}%
\end{equation}
where $\mathbf{u}\in\mathbb{R}^{2n}$,%
\begin{equation}
\mathbf{\hat{R}}^{n}=\left[  \left.
\begin{array}
[c]{ccc}%
\hat{x}_{1} & \cdots & \hat{x}_{n}%
\end{array}
\right\vert
\begin{array}
[c]{ccc}%
\hat{p}_{1} & \cdots & \hat{p}_{n}%
\end{array}
\right]  ^{T},
\end{equation}
and $\cdot$ is the inner product.

We can phrase continuous-variable quantum error correction theory in terms of
the operators resulting from the maps $\mathbf{D}$ and $\mathbf{M}$ or in
terms of the real vectors that result from the inverse maps $\mathbf{D}^{-1}$
and $\mathbf{M}^{-1}$. Both ways prove to be useful.

\section{Canonical Entanglement-Assisted Operator Continuous-Variable Code}

We begin our development of continuous-variable entanglement-assisted operator
coding by introducing a canonical code. This canonical code actively and
passively corrects for errors $\mathbf{D}\left(  \mathbf{u}\right)  $ in a
canonical error set $S_{0}$ where $\mathbf{u}\in S_{0}\subset\mathbb{R}^{2n}$.

Suppose Alice wishes to protect a $k$-mode quantum state $\left\vert
\psi\right\rangle $:%
\begin{equation}
\left\vert \psi\right\rangle =%
{\textstyle\idotsint}
dx_{1}\cdots dx_{k}\ \psi\left(  x_{1},\ldots,x_{k}\right)  \ \left\vert
x_{1}\right\rangle \cdots\left\vert x_{k}\right\rangle .
\end{equation}
Alice and Bob possess $c$ sets of infinitely-squeezed, perfectly entangled
states $\left\vert \Phi\right\rangle ^{\otimes c}$ where%
\begin{equation}
\left\vert \Phi\right\rangle \equiv\left(  \int\ dx\ \left\vert x\right\rangle
^{A}\left\vert x\right\rangle ^{B}\right)  /\sqrt{\pi}.
\end{equation}
The state $\left\vert \Phi\right\rangle $ is a zero-valued eigenstate of the
relative position observable $\hat{x}_{A}-\hat{x}_{B}$\ and total momentum
observable $\hat{p}_{A}+\hat{p}_{B}$. Alice possesses $l=n-k-c-r$ ancilla
modes initialized to infinitely-squeezed zero-position eigenstates of the
position observables $\hat{x}_{k+1},\ldots,\hat{x}_{k+l}$: $\left\vert
\mathbf{0}\right\rangle =\left\vert 0\right\rangle ^{\otimes l}$. Alice also
possesses an arbitrary mixed quantum state $\sigma$\ over $r$ modes. These $r$
modes are the gauge modes. She encodes the state $\left\vert \psi\right\rangle
$ with the canonical isometric encoder as follows:%
\begin{equation}
U_{0}:\left\vert \psi\right\rangle \left\langle \psi\right\vert \rightarrow
\left\vert \psi\right\rangle \left\langle \psi\right\vert \otimes\left\vert
\mathbf{0}\right\rangle \left\langle \mathbf{0}\right\vert \otimes
\sigma\otimes\left\vert \Phi\right\rangle \left\langle \Phi\right\vert .
\label{eq:canon_encode}%
\end{equation}
The canonical encoder merely appends the $l$ ancilla modes, $r$ gauge modes,
and $c$ entangled modes to the $k$ information modes.

Continuous-variable\ errors are equivalent to translations in position and
kicks in momentum \cite{prl1998braunstein_error,PhysRevA.64.012310}. The
canonical code corrects the error set%
\begin{equation}
S_{0}=\left\{
\begin{array}
[c]{c}%
\left(  \alpha\left(  \mathbf{a},\mathbf{a}_{1},\mathbf{a}_{2}\right)
,\mathbf{b,\mathbf{c},a}_{2}|\beta\left(  \mathbf{a},\mathbf{a}_{1}%
,\mathbf{a}_{2}\right)  ,\mathbf{a},\mathbf{d},\mathbf{a}_{1}\right) \\
:\mathbf{b,a}\in\mathbb{R}^{l},\ \ \mathbf{c,d}\in\mathbb{R}^{r}%
,\ \ \mathbf{a}_{1}\mathbf{,a}_{2}\in\mathbb{R}^{c}%
\end{array}
\right\}  ,
\end{equation}
for any known functions $\alpha,\beta:\mathbb{R}^{l}\times\mathbb{R}^{c}%
\times\mathbb{R}^{c}\rightarrow\mathbb{R}^{k}$. Consider an arbitrary error
$\mathbf{D}\left(  \mathbf{u}\right)  $ where%
\begin{equation}
\mathbf{u}=\left(  \alpha\left(  \mathbf{a},\mathbf{a}_{1},\mathbf{a}%
_{2}\right)  ,\mathbf{b,\mathbf{c},a}_{2}|\beta\left(  \mathbf{a}%
,\mathbf{a}_{1},\mathbf{a}_{2}\right)  ,\mathbf{a},\mathbf{d},\mathbf{a}%
_{1}\right)  . \label{eq:cv_error}%
\end{equation}
Suppose an error $\mathbf{D}\left(  \mathbf{u}\right)  $ occurs. State
$\left\vert \psi\right\rangle \left\langle \psi\right\vert \otimes\left\vert
\mathbf{0}\right\rangle \left\langle \mathbf{0}\right\vert \otimes
\sigma\otimes\left\vert \Phi\right\rangle \left\langle \Phi\right\vert $
becomes as follows (up to a global phase)%
\begin{multline}
\mathbf{Z}\left(  \alpha\right)  \mathbf{X}\left(  \beta\right)  \left\vert
\psi\right\rangle \left\langle \psi\right\vert \mathbf{X}\left(
-\beta\right)  \mathbf{Z}\left(  -\alpha\right)  \otimes\left\vert
\mathbf{a}\right\rangle \left\langle \mathbf{a}\right\vert \otimes
\sigma^{\prime}\otimes\\
\left\vert \mathbf{a}_{1},\mathbf{a}_{2}\right\rangle \left\langle
\mathbf{a}_{1},\mathbf{a}_{2}\right\vert ,
\end{multline}
where $\left\vert \mathbf{a}\right\rangle =\mathbf{X}\left(  \mathbf{a}%
\right)  \left\vert \mathbf{0}\right\rangle $, $\left\vert \mathbf{a}%
_{1},\mathbf{a}_{2}\right\rangle =\mathbf{X}\left(  \mathbf{a}_{1}\right)
\mathbf{Z}\left(  \mathbf{a}_{2}\right)  \left\vert \Phi\right\rangle
^{\otimes c}$, and%
\begin{equation}
\sigma^{\prime}=\mathbf{Z}\left(  \mathbf{c}\right)  \mathbf{X}\left(
\mathbf{d}\right)  \sigma\mathbf{X}\left(  -\mathbf{d}\right)  \mathbf{Z}%
\left(  -\mathbf{c}\right)  .
\end{equation}
Bob measures the position observables of the ancillas $\left\vert
\mathbf{a}\right\rangle $ and the relative position and total momentum
observables of the state $\left\vert \mathbf{a}_{1},\mathbf{a}_{2}%
\right\rangle $. He obtains a reduced error syndrome $\mathbf{r}=\left(
\mathbf{a,a}_{1},\mathbf{a}_{2}\right)  $. The reduced error syndrome
specifies the error up to an irrelevant value of $\mathbf{b}$, $\mathbf{c}$,
and $\mathbf{d}$ in (\ref{eq:cv_error}). The $\mathbf{b}$ errors are
irrelevant because the ancilla modes absorb these errors (the ancillas are
eigenstates of these error operators, and hence are unaffected by them.) The
$\mathbf{c}$ and $\mathbf{d}$ errors are irrelevant because they affect the
gauge modes only. Bob reverses the error $\mathbf{D}\left(  \mathbf{u}\right)
$ by applying the map $\mathbf{D}\left(  -\mathbf{u}^{\prime}\right)  $ where%
\begin{equation}
\mathbf{u}^{\prime}=\left(  \alpha\left(  \mathbf{a},\mathbf{a}_{1}%
,\mathbf{a}_{2}\right)  ,\mathbf{0},\mathbf{0,a}_{2}|\beta\left(
\mathbf{a},\mathbf{a}_{1},\mathbf{a}_{2}\right)  ,\mathbf{a},\mathbf{0}%
,\mathbf{a}_{1}\right)  . \label{eq:reverse_error}%
\end{equation}
This operation reverses the error because the states $\left\vert
\psi\right\rangle \left\langle \psi\right\vert \otimes\sigma$ and $\left\vert
\psi\right\rangle \left\langle \psi\right\vert \otimes\sigma^{\prime}$ differ
by a gauge operation only and thus possess the same quantum information.

The canonical code is a simple example of a continuous-variable
entanglement-assisted operator code, but it illustrates all of the principles
that are at work in the operation of an entanglement-assisted operator code.

\section{Parity-Check Matrix for the Canonical Code}

We now illustrate how the code operates in the Heisenberg picture by using a
parity check matrix. A parity check matrix $F_{0}$ characterizes the operators
that Bob measures:%
\begin{align}
F_{0}  &  =\left[  \left.
\begin{array}
[c]{ccccc}%
0 & I & 0 & 0 & 0\\
0 & 0 & 0 & I & -I\\
0 & 0 & 0 & 0 & 0
\end{array}
\right\vert
\begin{array}
[c]{ccccc}%
0 & 0 & 0 & 0 & 0\\
0 & 0 & 0 & 0 & 0\\
0 & 0 & 0 & I & I
\end{array}
\right]
\begin{array}
[c]{l}%
\}\ \ l\\
\}\ \ c\\
\}\ \ c
\end{array}
\\
&  =\left[  \left.  F_{Z0}\ \ \
\begin{array}
[c]{c}%
0\\
-I\\
0
\end{array}
\right\vert F_{X0}\ \ \
\begin{array}
[c]{c}%
0\\
0\\
I
\end{array}
\right]
\begin{array}
[c]{l}%
\}\ \ l\\
\}\ \ c\\
\}\ \ c
\end{array}
,
\end{align}
where%
\begin{equation}
F_{Z0}=\left[
\begin{array}
[c]{cccc}%
0 & I & 0 & 0\\
0 & 0 & 0 & I\\
0 & 0 & 0 & 0
\end{array}
\right]  ,\ \ F_{X0}=\left[
\begin{array}
[c]{cccc}%
0 & 0 & 0 & 0\\
0 & 0 & 0 & 0\\
0 & 0 & 0 & I
\end{array}
\right]  .
\end{equation}
These measurements diagnose errors on the modes that Alice sends over the
noisy channel. The first column of zeros in $F_{Z0}$ and $F_{X0}$ has $k$
entries and corresponds to the $k$ information modes. The third column of
zeros in $F_{Z0}$ and $F_{X0}$ has $r$ entries and corresponds to the $r$
gauge modes. The entries in $F_{Z0}$ and $F_{X0}$ correspond to the modes that
Alice initially possesses and the last $c$ columns in $F_{0}$ to the right of
$F_{Z0}$ and $F_{X0}$ correspond to the modes that Bob initially possesses.
Noise does not affect these $c$\ modes on Bob's side because they are on the
receiving end of the channel. The map $\mathbf{M}$ determines the observables
that Bob measures to learn about the errors. Each row $\mathbf{f}$\ of $F_{0}$
corresponds to an element of the set
\begin{equation}
\mathcal{M}_{0}\equiv\left\{  \mathbf{M}\left(  \mathbf{f}\right)
:\mathbf{f}\text{ is a row of }F_{0}\right\}  .
\end{equation}
Therefore, the first $l$ rows of $F_{0}$ correspond to the $l$ position
observables and the last $2c$ rows of $F_{0}$ correspond to the relative
position and total momentum observables. Matrix $F_{0}$\ thus gives another
way of describing the measurements performed in the canonical code. Bob
measures the observables in $\mathcal{M}_{0}$\ to learn about the error
without disturbing the encoded state.

The following gauge matrix $G_{0}$ characterizes the errors that the code
passively corrects due to the presence of gauge modes:%
\begin{align}
G_{0}  &  =\left[  \left.
\begin{array}
[c]{ccccc}%
0 & 0 & I & 0 & 0\\
0 & 0 & 0 & 0 & 0
\end{array}
\right\vert
\begin{array}
[c]{ccccc}%
0 & 0 & 0 & 0 & 0\\
0 & 0 & I & 0 & 0
\end{array}
\right]
\begin{array}
[c]{l}%
\}\ \ r\\
\}\ \ r
\end{array}
\\
&  =\left[  \left.  G_{Z0}\ \ \
\begin{array}
[c]{c}%
0\\
0
\end{array}
\right\vert G_{X0}\ \ \
\begin{array}
[c]{c}%
0\\
0
\end{array}
\right]
\begin{array}
[c]{l}%
\}\ \ r\\
\}\ \ r
\end{array}
.
\end{align}
The entries in $G_{0}$ form a basis for passively correctable errors.
Therefore, the code passively corrects errors in the following set:%
\begin{equation}
\mathcal{G}_{0}\equiv\left\{  \mathbf{D}\left(  \mathbf{g}\right)
:\mathbf{g}\in\text{rowspace}\left(  G_{0}\right)  \right\}  .
\end{equation}
This passive correction of errors is the additional benefit of including gauge
modes in our codes. This incorporation of gauge modes may be able to help in
correcting the small errors that plague continuous-variable quantum
information systems.

The canonical code can correct an error set $\mathcal{E}_{0}$ that consists of
all pairs of errors obeying the following condition: $\forall\ \mathbf{D}%
\left(  \mathbf{e}\right)  ,\mathbf{D}\left(  \mathbf{e}^{\prime}\right)
\in\mathcal{E}_{0}$ with $\mathbf{e}\neq\mathbf{e}^{\prime}$ either
\begin{equation}
\mathbf{e}-\mathbf{e}^{\prime}\notin\ \left(  \text{rowspace}\left(
F_{0,I}\right)  \oplus\text{rowspace}\left(  F_{0,E}\right)  \right)  ^{\perp
},
\end{equation}
or%
\begin{equation}
\mathbf{e}-\mathbf{e}^{\prime}\in\text{rowspace}\left(  F_{0,I}\right)
\oplus\text{rowspace}\left(  G_{0}\right)  ,
\end{equation}
where%
\begin{align}
F_{0,I}  &  =\left[  \left.
\begin{array}
[c]{cccc}%
0 & I & 0 & 0
\end{array}
\right\vert
\begin{array}
[c]{cccc}%
0 & 0 & 0 & 0
\end{array}
\right]  ,\\
F_{0,E}  &  =\left[  \left.
\begin{array}
[c]{cccc}%
0 & 0 & 0 & I\\
0 & 0 & 0 & 0
\end{array}
\right\vert
\begin{array}
[c]{cccc}%
0 & 0 & 0 & 0\\
0 & 0 & 0 & I
\end{array}
\right]  ,
\end{align}
and $\perp$ denotes the symplectic dual \cite{pra2007wildeEA}.

\section{General Entanglement-Assisted Operator Codes}

The relation between the canonical entanglement-assisted operator code and an
arbitrary one is similar to the relation found in Ref. \cite{pra2007wildeEA}.
Alice can perform the encoding of an arbitrary code with a local unitary $U$.
This local unitary $U$\ preserves operators in the phase-free Heisenberg-Weyl
group under conjugation \cite{pra2007wildeEA}\ and relates the canonical code
to an arbitrary one. An equivalent representation of $U$ is with a symplectic
matrix $\Upsilon$ that operates on the real vectors that result from the
inverse maps $\mathbf{D}^{-1}$ and $\mathbf{M}^{-1}$. The former statement is
equivalent to Theorem~2 from Ref.~\cite{pra2007wildeEA}.

A local unitary $U$ operating on the first $n$ modes relates the canonical
code to a general one. In the Heisenberg picture, the symplectic matrix
$\Upsilon$ is a $\left(  2n\times2n\right)  $-dimensional matrix that takes
the canonical parity check matrix $F_{0}$\ to a general check matrix $F$ and
the gauge matrix $G_{0}$\ to a general gauge matrix $G$. The symplectic matrix
$\Upsilon$ then performs the following transformation:%
\begin{align}
\left[
\begin{array}
[c]{cc}%
F_{Z0} & F_{X0}%
\end{array}
\right]  \Upsilon^{T}  &  =\left[
\begin{array}
[c]{cc}%
F_{Z} & F_{X}%
\end{array}
\right]  ,\label{eq:gen-error-cond-mat}\\
\left[
\begin{array}
[c]{cc}%
G_{Z0} & G_{X0}%
\end{array}
\right]  \Upsilon^{T}  &  =\left[
\begin{array}
[c]{cc}%
G_{Z} & G_{X}%
\end{array}
\right]  .
\end{align}
The parity check matrix $F$ for a general code has the following form:%
\begin{equation}
F=\left[  \left.  F_{Z}\ \ \
\begin{array}
[c]{c}%
0\\
-I\\
0
\end{array}
\right\vert F_{X}\ \ \
\begin{array}
[c]{c}%
0\\
0\\
I
\end{array}
\right]
\begin{array}
[c]{l}%
\}\ \ l\\
\}\ \ c\\
\}\ \ c
\end{array}
,
\end{equation}
and the gauge matrix $G$ has the following form:%
\begin{equation}
G=\left[  \left.  G_{Z}\ \ \
\begin{array}
[c]{c}%
0\\
0
\end{array}
\right\vert G_{X}\ \ \
\begin{array}
[c]{c}%
0\\
0
\end{array}
\right]
\begin{array}
[c]{l}%
\}\ \ r\\
\}\ \ r
\end{array}
.
\end{equation}
Bob measures the observables in the set%
\begin{equation}
\mathcal{M}\equiv\left\{  \mathbf{M}\left(  \mathbf{f}\right)  :\mathbf{f}%
\text{ is a row of }F\right\}  ,
\end{equation}
to diagnose and correct for errors. The code has passive protection against
errors in the following set:%
\begin{equation}
\mathcal{G}\equiv\left\{  \mathbf{D}\left(  \mathbf{g}\right)  :\mathbf{g}%
\in\text{rowspace}\left(  G\right)  \right\}  .
\end{equation}

The error-correcting conditions for our continuous-variable
entanglement-assisted operator codes include those given in
Ref.~\cite{pra2007wildeEA}. These codes also have some additional passive
error-correcting capability due to the inclusion of gauge modes. Our codes can
correct for all errors satisfying the conditions in Ref.~\cite{pra2007wildeEA}
and all errors $\mathbf{D}\left(  \mathbf{u}\right)  \in\mathcal{G}$. A
general code can correct an error set $\mathcal{E}$ that consists of all pairs
of errors obeying the following condition: $\forall\ \mathbf{D}\left(
\mathbf{e}\right)  ,\mathbf{D}\left(  \mathbf{e}\right)  ^{\prime}%
\in\mathcal{E}$ with $\mathbf{e}\neq\mathbf{e}^{\prime}$ either
\begin{equation}
\mathbf{e}-\mathbf{e}^{\prime}\notin\ \left(  \text{rowspace}\left(
F_{I}\right)  \oplus\text{rowspace}\left(  F_{E}\right)  \right)  ^{\perp},
\end{equation}
or%
\begin{equation}
\mathbf{e}-\mathbf{e}^{\prime}\in\text{rowspace}\left(  F_{I}\right)
\oplus\text{rowspace}\left(  G\right)  ,
\end{equation}
where $F_{I}$ consists of the first $l$ rows of the matrix on the RHS\ of
(\ref{eq:gen-error-cond-mat}) and $F_{E}$ consists of the last $2c$ rows of
the matrix on the RHS\ of (\ref{eq:gen-error-cond-mat}) and $\perp$ denotes
the symplectic dual \cite{pra2007wildeEA}.

\section{Example}

We present an example of a continuous-variable entanglement-assisted operator
code. This code is a straightforward extension of the entanglement-assisted
Bacon-Shor code~from Ref.~\cite{hsieh:062313}. We employ the method that
Barnes suggested in Ref.~\cite{arx04barnes} that takes the stabilizer matrix
for a discrete code and replaces \textquotedblleft1\textquotedblright\ entries
with a \textquotedblleft1\textquotedblright\ or \textquotedblleft%
--1\textquotedblright\ to make the symplectic product between rows be equal to
one or zero. Our example encodes one information mode with the help of one set
of entangled modes, four ancilla modes, and two gauge modes.

Its initial unencoded check matrix is as follows:%
\[
F_{0}=\left[  \left.
\begin{array}
[c]{cccccccc}%
1 & 0 & 0 & 0 & 0 & 0 & 0 & 0\\
0 & 1 & 0 & 0 & 0 & 0 & 0 & 0\\
0 & 0 & 1 & 0 & 0 & 0 & 0 & 0\\
0 & 0 & 0 & 1 & 0 & 0 & 0 & 0\\
0 & 0 & 0 & 0 & 0 & 0 & 0 & 0\\
0 & 0 & 0 & 0 & 1 & 0 & 0 & 0
\end{array}
\right\vert
\begin{array}
[c]{cccccccc}%
0 & 0 & 0 & 0 & 0 & 0 & 0 & 0\\
0 & 0 & 0 & 0 & 0 & 0 & 0 & 0\\
0 & 0 & 0 & 0 & 0 & 0 & 0 & 0\\
0 & 0 & 0 & 0 & 0 & 0 & 0 & 0\\
0 & 0 & 0 & 1 & 0 & 0 & 0 & 0\\
0 & 0 & 0 & 0 & 0 & 0 & 0 & 0
\end{array}
\right]  .
\]
Rows four and five of the above matrix correspond to half of an entangled mode
and the other rows correspond to ancilla modes. The initial matrix for the
gauge operators is as follows:%
\[
G_{0}=\left[  \left.
\begin{array}
[c]{cccccccc}%
0 & 0 & 0 & 0 & 0 & 1 & 0 & 0\\
0 & 0 & 0 & 0 & 0 & 0 & 0 & 0\\
0 & 0 & 0 & 0 & 0 & 0 & 1 & 0\\
0 & 0 & 0 & 0 & 0 & 0 & 0 & 0
\end{array}
\right\vert
\begin{array}
[c]{cccccccc}%
0 & 0 & 0 & 0 & 0 & 0 & 0 & 0\\
0 & 0 & 0 & 0 & 0 & 1 & 0 & 0\\
0 & 0 & 0 & 0 & 0 & 0 & 0 & 0\\
0 & 0 & 0 & 0 & 0 & 0 & 1 & 0
\end{array}
\right]  .
\]
The information mode corresponds to the last column of each of the above submatrices.

A linear-optical encoding operation (described in the next
section)\ transforms the unencoded state to the encoded state. The check
matrix corresponding to the encoded state is as follows:%
\[
F=\left[
\begin{array}
[c]{c}%
F_{Z}%
\end{array}
\left\vert
\begin{array}
[c]{c}%
F_{X}%
\end{array}
\right.  \right]  ,
\]
where%
\begin{align*}
F_{Z}  &  =\left[
\begin{array}
[c]{cccccccc}%
1 & -1 & 0 & 1 & -1 & 0 & 0 & 0\\
1 & 0 & -1 & 1 & 0 & -1 & 0 & 0\\
0 & 0 & 0 & 0 & 0 & 0 & 1 & -1\\
0 & 0 & 0 & 0 & 0 & 0 & 0 & 0\\
0 & 0 & 0 & 0 & 0 & 0 & 0 & 1\\
0 & 0 & 0 & 0 & 0 & 0 & 0 & 0
\end{array}
\right]  ,\\
F_{X}  &  =\left[
\begin{array}
[c]{cccccccc}%
0 & 0 & 0 & 0 & 0 & 0 & 0 & 0\\
0 & 0 & 0 & 0 & 0 & 0 & 0 & 0\\
0 & 0 & 0 & 0 & 0 & 0 & 0 & 0\\
1 & 1 & 1 & 0 & 0 & 0 & -1 & -1\\
0 & 0 & 0 & 0 & 0 & 0 & 0 & 0\\
1 & 1 & 1 & -1 & -1 & -1 & 0 & 0
\end{array}
\right]  .
\end{align*}
The matrix corresponding to the gauge operators is as follows:%
\[
G=\left[  \left.
\begin{array}
[c]{cccccccc}%
0 & 0 & 0 & 0 & 0 & 0 & 0 & 0\\
1 & -1 & 0 & 0 & 0 & 0 & 0 & 0\\
0 & 0 & 0 & 1 & 0 & -1 & 0 & 0\\
0 & 0 & 0 & 0 & 0 & 0 & 0 & 0
\end{array}
\right\vert
\begin{array}
[c]{cccccccc}%
0 & 1 & 0 & 0 & -1 & 0 & 0 & 0\\
0 & 0 & 0 & 0 & 0 & 0 & 0 & 0\\
0 & 0 & 0 & 0 & 0 & 0 & 0 & 0\\
0 & 0 & 1 & 0 & 0 & -1 & 0 & 0
\end{array}
\right]  .
\]
The code passively corrects error in the above gauge group. The receiver Bob
measures the operators corresponding to rows one, two, three, and six in the
matrix $F$. Bob combines his half of the entangled mode and measures operators
corresponding to the following augmented version of rows four and five of $F$:%
\[
F_{\text{aug}}=\left[  \left.
\begin{array}
[c]{cccccccc}%
0 & 0 & 0 & 0 & 0 & 0 & 0 & 0\\
0 & 0 & 0 & 0 & 0 & 0 & 0 & 1
\end{array}%
\begin{array}
[c]{c}%
0\\
1
\end{array}
\right\vert
\begin{array}
[c]{cccccccc}%
1 & 1 & 1 & 0 & 0 & 0 & -1 & -1\\
0 & 0 & 0 & 0 & 0 & 0 & 0 & 0
\end{array}%
\begin{array}
[c]{c}%
1\\
0
\end{array}
\right]  .
\]
The code corrects an arbitrary single-mode error. This error-correcting
capability follows directly from the discrete-variable code's error-correcting properties.

\section{Encoding Circuit}

Two different algorithms exist for constructing a linear-optical encoding
circuit corresponding to the encoding unitary $U$
\cite{braunstein:055801,pra2007wildeEA}. The algorithm in
Ref.~\cite{braunstein:055801} uses the Bloch-Messiah transformation to
decompose a symplectic matrix into a sequence of passive optical
transformations, online squeezers, and passive optical transformations. One
can use the technique of Filip \textit{et al.} for implementing the online
squeezers. It is also possible to use the algorithm in
Ref.~\cite{pra2007wildeEA} for a linear-optical encoding circuit, but this
technique uses quantum nondemolition interactions and may be more difficult to
implement experimentally.

\section{Concluding Remarks}

Our EAQECCs are vulnerable to finite squeezing effects and inefficient
photodetectors for the same reasons as those in
Refs.~\cite{prl1998braunstein_error,pra2007wildeEA}. Our scheme works well if
the errors due to finite squeezing and inefficiencies in beamsplitters and
photodetectors are smaller than the actual errors.

M.M.W. acknowledges support from NSF\ Grants CCF-0545845\ and CCF-0448658, and
T.A.B. acknowledges support from NSF Grant CCF-0448658.

\bibliographystyle{apsrev}
\bibliography{eaoqec-cv}

\begin{thebibliography}{45}
\expandafter\ifx\csname natexlab\endcsname\relax\def\natexlab#1{#1}\fi
\expandafter\ifx\csname bibnamefont\endcsname\relax
  \def\bibnamefont#1{#1}\fi
\expandafter\ifx\csname bibfnamefont\endcsname\relax
  \def\bibfnamefont#1{#1}\fi
\expandafter\ifx\csname citenamefont\endcsname\relax
  \def\citenamefont#1{#1}\fi
\expandafter\ifx\csname url\endcsname\relax
  \def\url#1{\texttt{#1}}\fi
\expandafter\ifx\csname urlprefix\endcsname\relax\def\urlprefix{URL }\fi
\providecommand{\bibinfo}[2]{#2}
\providecommand{\eprint}[2][]{\url{#2}}

\bibitem[{\citenamefont{Shor}(1995)}]{PhysRevA.52.R2493}
\bibinfo{author}{\bibfnamefont{P.~W.} \bibnamefont{Shor}},
  \bibinfo{journal}{Phys. Rev. A} \textbf{\bibinfo{volume}{52}},
  \bibinfo{pages}{R2493} (\bibinfo{year}{1995}).

\bibitem[{\citenamefont{Steane}(1996)}]{PhysRevLett.77.793}
\bibinfo{author}{\bibfnamefont{A.~M.} \bibnamefont{Steane}},
  \bibinfo{journal}{Phys. Rev. Lett.} \textbf{\bibinfo{volume}{77}},
  \bibinfo{pages}{793} (\bibinfo{year}{1996}).

\bibitem[{\citenamefont{Calderbank and Shor}(1996)}]{PhysRevA.54.1098}
\bibinfo{author}{\bibfnamefont{A.~R.} \bibnamefont{Calderbank}}
  \bibnamefont{and} \bibinfo{author}{\bibfnamefont{P.~W.} \bibnamefont{Shor}},
  \bibinfo{journal}{Phys. Rev. A} \textbf{\bibinfo{volume}{54}},
  \bibinfo{pages}{1098} (\bibinfo{year}{1996}).

\bibitem[{\citenamefont{Gottesman}(1997)}]{thesis97gottesman}
\bibinfo{author}{\bibfnamefont{D.}~\bibnamefont{Gottesman}}, Ph.D. thesis,
  \bibinfo{school}{California Institue of Technology} (\bibinfo{year}{1997}).

\bibitem[{\citenamefont{Calderbank et~al.}(1997)\citenamefont{Calderbank,
  Rains, Shor, and Sloane}}]{PhysRevLett.78.405}
\bibinfo{author}{\bibfnamefont{A.~R.} \bibnamefont{Calderbank}},
  \bibinfo{author}{\bibfnamefont{E.~M.} \bibnamefont{Rains}},
  \bibinfo{author}{\bibfnamefont{P.~W.} \bibnamefont{Shor}}, \bibnamefont{and}
  \bibinfo{author}{\bibfnamefont{N.~J.~A.} \bibnamefont{Sloane}},
  \bibinfo{journal}{Phys. Rev. Lett.} \textbf{\bibinfo{volume}{78}},
  \bibinfo{pages}{405} (\bibinfo{year}{1997}).

\bibitem[{\citenamefont{Calderbank et~al.}(1998)\citenamefont{Calderbank,
  Rains, Shor, and Sloane}}]{ieee1998calderbank}
\bibinfo{author}{\bibfnamefont{A.}~\bibnamefont{Calderbank}},
  \bibinfo{author}{\bibfnamefont{E.}~\bibnamefont{Rains}},
  \bibinfo{author}{\bibfnamefont{P.}~\bibnamefont{Shor}}, \bibnamefont{and}
  \bibinfo{author}{\bibfnamefont{N.}~\bibnamefont{Sloane}},
  \bibinfo{journal}{IEEE Trans. Inf. Theory} \textbf{\bibinfo{volume}{44}},
  \bibinfo{pages}{1369} (\bibinfo{year}{1998}).

\bibitem[{\citenamefont{Zanardi and
  Rasetti}(1997{\natexlab{a}})}]{PhysRevLett.79.3306}
\bibinfo{author}{\bibfnamefont{P.}~\bibnamefont{Zanardi}} \bibnamefont{and}
  \bibinfo{author}{\bibfnamefont{M.}~\bibnamefont{Rasetti}},
  \bibinfo{journal}{Phys. Rev. Lett.} \textbf{\bibinfo{volume}{79}},
  \bibinfo{pages}{3306} (\bibinfo{year}{1997}{\natexlab{a}}).

\bibitem[{\citenamefont{Zanardi and
  Rasetti}(1997{\natexlab{b}})}]{mpl1997zanardi}
\bibinfo{author}{\bibfnamefont{P.}~\bibnamefont{Zanardi}} \bibnamefont{and}
  \bibinfo{author}{\bibfnamefont{M.}~\bibnamefont{Rasetti}},
  \bibinfo{journal}{Mod. Phys. Lett. B} \textbf{\bibinfo{volume}{11}},
  \bibinfo{pages}{1085} (\bibinfo{year}{1997}{\natexlab{b}}).

\bibitem[{\citenamefont{Lidar et~al.}(1998)\citenamefont{Lidar, Chuang, and
  Whaley}}]{PhysRevLett.81.2594}
\bibinfo{author}{\bibfnamefont{D.~A.} \bibnamefont{Lidar}},
  \bibinfo{author}{\bibfnamefont{I.~L.} \bibnamefont{Chuang}},
  \bibnamefont{and} \bibinfo{author}{\bibfnamefont{K.~B.}
  \bibnamefont{Whaley}}, \bibinfo{journal}{Phys. Rev. Lett.}
  \textbf{\bibinfo{volume}{81}}, \bibinfo{pages}{2594} (\bibinfo{year}{1998}).

\bibitem[{\citenamefont{Kribs et~al.}(2005)\citenamefont{Kribs, Laflamme, and
  Poulin}}]{kribs:180501}
\bibinfo{author}{\bibfnamefont{D.}~\bibnamefont{Kribs}},
  \bibinfo{author}{\bibfnamefont{R.}~\bibnamefont{Laflamme}}, \bibnamefont{and}
  \bibinfo{author}{\bibfnamefont{D.}~\bibnamefont{Poulin}},
  \bibinfo{journal}{Phys. Rev. Lett.} \textbf{\bibinfo{volume}{94}},
  \bibinfo{eid}{180501} (\bibinfo{year}{2005}).

\bibitem[{\citenamefont{Kribs et~al.}(2006)\citenamefont{Kribs, Laflamme,
  Poulin, and Lesosky}}]{qic2006kribs}
\bibinfo{author}{\bibfnamefont{D.~W.} \bibnamefont{Kribs}},
  \bibinfo{author}{\bibfnamefont{R.}~\bibnamefont{Laflamme}},
  \bibinfo{author}{\bibfnamefont{D.}~\bibnamefont{Poulin}}, \bibnamefont{and}
  \bibinfo{author}{\bibfnamefont{M.}~\bibnamefont{Lesosky}},
  \bibinfo{journal}{Quant. Inf. \& Comp.} \textbf{\bibinfo{volume}{6}},
  \bibinfo{pages}{383} (\bibinfo{year}{2006}).

\bibitem[{\citenamefont{Poulin}(2005)}]{poulin:230504}
\bibinfo{author}{\bibfnamefont{D.}~\bibnamefont{Poulin}},
  \bibinfo{journal}{Phys. Rev. Lett.} \textbf{\bibinfo{volume}{95}},
  \bibinfo{eid}{230504} (\bibinfo{year}{2005}).

\bibitem[{\citenamefont{Brun et~al.}(2007)\citenamefont{Brun, Devetak, and
  Hsieh}}]{isit2007brun}
\bibinfo{author}{\bibfnamefont{T.}~\bibnamefont{Brun}},
  \bibinfo{author}{\bibfnamefont{I.}~\bibnamefont{Devetak}}, \bibnamefont{and}
  \bibinfo{author}{\bibfnamefont{M.-H.} \bibnamefont{Hsieh}}, in
  \emph{\bibinfo{booktitle}{IEEE International Symposium on Information
  Theory}} (\bibinfo{year}{2007}), pp. \bibinfo{pages}{2101--2105}.

\bibitem[{\citenamefont{Hsieh et~al.}(2007)\citenamefont{Hsieh, Devetak, and
  Brun}}]{hsieh:062313}
\bibinfo{author}{\bibfnamefont{M.-H.} \bibnamefont{Hsieh}},
  \bibinfo{author}{\bibfnamefont{I.}~\bibnamefont{Devetak}}, \bibnamefont{and}
  \bibinfo{author}{\bibfnamefont{T.}~\bibnamefont{Brun}},
  \bibinfo{journal}{Phys. Rev. A} \textbf{\bibinfo{volume}{76}},
  \bibinfo{eid}{062313} (\bibinfo{year}{2007}).

\bibitem[{\citenamefont{Braunstein and Pati}(2003)}]{book2003braunstein}
\bibinfo{editor}{\bibfnamefont{S.~L.} \bibnamefont{Braunstein}}
  \bibnamefont{and} \bibinfo{editor}{\bibfnamefont{A.}~\bibnamefont{Pati}},
  eds., \emph{\bibinfo{title}{Quantum Information with Continuous Variables}}
  (\bibinfo{publisher}{Springer}, \bibinfo{year}{2003}).

\bibitem[{\citenamefont{Braunstein and van Loock}(2005)}]{revmod2005braunstein}
\bibinfo{author}{\bibfnamefont{S.~L.} \bibnamefont{Braunstein}}
  \bibnamefont{and} \bibinfo{author}{\bibfnamefont{P.}~\bibnamefont{van
  Loock}}, \bibinfo{journal}{Rev. Mod. Phys.} \textbf{\bibinfo{volume}{77}},
  \bibinfo{pages}{513} (\bibinfo{year}{2005}).

\bibitem[{\citenamefont{Furusawa et~al.}(1998)\citenamefont{Furusawa, Sørensen,
  Braunstein, Fuchs, Kimble, and Polzik}}]{science1998furusawa}
\bibinfo{author}{\bibfnamefont{A.}~\bibnamefont{Furusawa}},
  \bibinfo{author}{\bibfnamefont{J.~L.} \bibnamefont{Sørensen}},
  \bibinfo{author}{\bibfnamefont{S.~L.} \bibnamefont{Braunstein}},
  \bibinfo{author}{\bibfnamefont{C.~A.} \bibnamefont{Fuchs}},
  \bibinfo{author}{\bibfnamefont{H.~J.} \bibnamefont{Kimble}},
  \bibnamefont{and} \bibinfo{author}{\bibfnamefont{E.~S.}
  \bibnamefont{Polzik}}, \bibinfo{journal}{Science}
  \textbf{\bibinfo{volume}{282}}, \bibinfo{pages}{706} (\bibinfo{year}{1998}).

\bibitem[{\citenamefont{Mizuno et~al.}(2005)\citenamefont{Mizuno, Wakui,
  Furusawa, and Sasaki}}]{mizuno:012304}
\bibinfo{author}{\bibfnamefont{J.}~\bibnamefont{Mizuno}},
  \bibinfo{author}{\bibfnamefont{K.}~\bibnamefont{Wakui}},
  \bibinfo{author}{\bibfnamefont{A.}~\bibnamefont{Furusawa}}, \bibnamefont{and}
  \bibinfo{author}{\bibfnamefont{M.}~\bibnamefont{Sasaki}},
  \bibinfo{journal}{Phys. Rev. A} \textbf{\bibinfo{volume}{71}},
  \bibinfo{eid}{012304} (\bibinfo{year}{2005}).

\bibitem[{\citenamefont{Li et~al.}(2002)\citenamefont{Li, Pan, Jing, Zhang,
  Xie, and Peng}}]{PhysRevLett.88.047904}
\bibinfo{author}{\bibfnamefont{X.}~\bibnamefont{Li}},
  \bibinfo{author}{\bibfnamefont{Q.}~\bibnamefont{Pan}},
  \bibinfo{author}{\bibfnamefont{J.}~\bibnamefont{Jing}},
  \bibinfo{author}{\bibfnamefont{J.}~\bibnamefont{Zhang}},
  \bibinfo{author}{\bibfnamefont{C.}~\bibnamefont{Xie}}, \bibnamefont{and}
  \bibinfo{author}{\bibfnamefont{K.}~\bibnamefont{Peng}},
  \bibinfo{journal}{Phys. Rev. Lett.} \textbf{\bibinfo{volume}{88}},
  \bibinfo{pages}{047904} (\bibinfo{year}{2002}).

\bibitem[{\citenamefont{Vaidman}(1994)}]{PhysRevA.49.1473}
\bibinfo{author}{\bibfnamefont{L.}~\bibnamefont{Vaidman}},
  \bibinfo{journal}{Phys. Rev. A} \textbf{\bibinfo{volume}{49}},
  \bibinfo{pages}{1473} (\bibinfo{year}{1994}).

\bibitem[{\citenamefont{Braunstein and Kimble}(1998)}]{prl1998braunstein}
\bibinfo{author}{\bibfnamefont{S.~L.} \bibnamefont{Braunstein}}
  \bibnamefont{and} \bibinfo{author}{\bibfnamefont{H.~J.}
  \bibnamefont{Kimble}}, \bibinfo{journal}{Phys. Rev. Lett.}
  \textbf{\bibinfo{volume}{80}}, \bibinfo{pages}{869} (\bibinfo{year}{1998}).

\bibitem[{\citenamefont{Ban}(1999)}]{optb1999ban}
\bibinfo{author}{\bibfnamefont{M.}~\bibnamefont{Ban}}, \bibinfo{journal}{J.
  Opt. B: Quantum Semiclass. Opt.} \textbf{\bibinfo{volume}{1}},
  \bibinfo{pages}{L1} (\bibinfo{year}{1999}).

\bibitem[{\citenamefont{Braunstein and Kimble}(2000)}]{pra2000braunstein}
\bibinfo{author}{\bibfnamefont{S.~L.} \bibnamefont{Braunstein}}
  \bibnamefont{and} \bibinfo{author}{\bibfnamefont{H.~J.}
  \bibnamefont{Kimble}}, \bibinfo{journal}{Phys. Rev. A}
  \textbf{\bibinfo{volume}{61}}, \bibinfo{pages}{042302}
  (\bibinfo{year}{2000}).

\bibitem[{\citenamefont{Filip et~al.}(2005)\citenamefont{Filip, Marek, and
  Andersen}}]{filip:042308}
\bibinfo{author}{\bibfnamefont{R.}~\bibnamefont{Filip}},
  \bibinfo{author}{\bibfnamefont{P.}~\bibnamefont{Marek}}, \bibnamefont{and}
  \bibinfo{author}{\bibfnamefont{U.~L.} \bibnamefont{Andersen}},
  \bibinfo{journal}{Phys. Rev. A} \textbf{\bibinfo{volume}{71}},
  \bibinfo{eid}{042308} (\bibinfo{year}{2005}).

\bibitem[{\citenamefont{Yoshikawa
  et~al.}(2007{\natexlab{a}})\citenamefont{Yoshikawa, Hayashi, Akiyama, Takei,
  Huck, Andersen, and Furusawa}}]{arx2007furusawa}
\bibinfo{author}{\bibfnamefont{J.-I.} \bibnamefont{Yoshikawa}},
  \bibinfo{author}{\bibfnamefont{T.}~\bibnamefont{Hayashi}},
  \bibinfo{author}{\bibfnamefont{T.}~\bibnamefont{Akiyama}},
  \bibinfo{author}{\bibfnamefont{N.}~\bibnamefont{Takei}},
  \bibinfo{author}{\bibfnamefont{A.}~\bibnamefont{Huck}},
  \bibinfo{author}{\bibfnamefont{U.~L.} \bibnamefont{Andersen}},
  \bibnamefont{and} \bibinfo{author}{\bibfnamefont{A.}~\bibnamefont{Furusawa}},
  \bibinfo{journal}{arXiv:quant-ph/0702049}
  (\bibinfo{year}{2007}{\natexlab{a}}).

\bibitem[{\citenamefont{Yoshikawa
  et~al.}(2007{\natexlab{b}})\citenamefont{Yoshikawa, Huck, Andersen, and
  Furusawa}}]{conf2007furusawa}
\bibinfo{author}{\bibfnamefont{J.}~\bibnamefont{Yoshikawa}},
  \bibinfo{author}{\bibfnamefont{A.}~\bibnamefont{Huck}},
  \bibinfo{author}{\bibfnamefont{U.~L.} \bibnamefont{Andersen}},
  \bibnamefont{and} \bibinfo{author}{\bibfnamefont{A.}~\bibnamefont{Furusawa}},
  in \emph{\bibinfo{booktitle}{Spring meeting of Physical Society of Japan,
  19aXG-2}} (\bibinfo{year}{2007}{\natexlab{b}}).

\bibitem[{\citenamefont{Wilde et~al.}(2007{\natexlab{a}})\citenamefont{Wilde,
  Krovi, and Brun}}]{pra2007wildeEA}
\bibinfo{author}{\bibfnamefont{M.~M.} \bibnamefont{Wilde}},
  \bibinfo{author}{\bibfnamefont{H.}~\bibnamefont{Krovi}}, \bibnamefont{and}
  \bibinfo{author}{\bibfnamefont{T.~A.} \bibnamefont{Brun}},
  \bibinfo{journal}{Phys. Rev. A} \textbf{\bibinfo{volume}{76}},
  \bibinfo{pages}{052308} (\bibinfo{year}{2007}{\natexlab{a}}).

\bibitem[{\citenamefont{Wilde et~al.}(2007{\natexlab{b}})\citenamefont{Wilde,
  Krovi, and Brun}}]{wilde:060303}
\bibinfo{author}{\bibfnamefont{M.~M.} \bibnamefont{Wilde}},
  \bibinfo{author}{\bibfnamefont{H.}~\bibnamefont{Krovi}}, \bibnamefont{and}
  \bibinfo{author}{\bibfnamefont{T.~A.} \bibnamefont{Brun}},
  \bibinfo{journal}{Phys. Rev. A} \textbf{\bibinfo{volume}{75}},
  \bibinfo{eid}{060303} (\bibinfo{year}{2007}{\natexlab{b}}).

\bibitem[{\citenamefont{Wilde et~al.}(2007{\natexlab{c}})\citenamefont{Wilde,
  Brun, Dowling, and Lee}}]{arx2007wildeCC}
\bibinfo{author}{\bibfnamefont{M.~M.} \bibnamefont{Wilde}},
  \bibinfo{author}{\bibfnamefont{T.~A.} \bibnamefont{Brun}},
  \bibinfo{author}{\bibfnamefont{J.~P.} \bibnamefont{Dowling}},
  \bibnamefont{and} \bibinfo{author}{\bibfnamefont{H.}~\bibnamefont{Lee}},
  \bibinfo{journal}{Phys. Rev. A} \textbf{\bibinfo{volume}{77}},
  \bibinfo{pages}{022321} (\bibinfo{year}{2007}{\natexlab{c}}).

\bibitem[{\citenamefont{Braunstein}(1998{\natexlab{a}})}]{prl1998braunstein_er%
ror}
\bibinfo{author}{\bibfnamefont{S.~L.} \bibnamefont{Braunstein}},
  \bibinfo{journal}{Phys. Rev. Lett.} \textbf{\bibinfo{volume}{80}},
  \bibinfo{pages}{4084} (\bibinfo{year}{1998}{\natexlab{a}}).

\bibitem[{\citenamefont{Lloyd and Slotine}(1998)}]{PhysRevLett.80.4088}
\bibinfo{author}{\bibfnamefont{S.}~\bibnamefont{Lloyd}} \bibnamefont{and}
  \bibinfo{author}{\bibfnamefont{J.-J.~E.} \bibnamefont{Slotine}},
  \bibinfo{journal}{Phys. Rev. Lett.} \textbf{\bibinfo{volume}{80}},
  \bibinfo{pages}{4088} (\bibinfo{year}{1998}).

\bibitem[{\citenamefont{Braunstein}(1998{\natexlab{b}})}]{nat1998braun}
\bibinfo{author}{\bibfnamefont{S.~L.} \bibnamefont{Braunstein}},
  \bibinfo{journal}{Nature} \textbf{\bibinfo{volume}{394}}, \bibinfo{pages}{47}
  (\bibinfo{year}{1998}{\natexlab{b}}).

\bibitem[{\citenamefont{Niset et~al.}(2007)\citenamefont{Niset, Andersen, and
  Cerf}}]{arx2007niset}
\bibinfo{author}{\bibfnamefont{J.}~\bibnamefont{Niset}},
  \bibinfo{author}{\bibfnamefont{U.~L.} \bibnamefont{Andersen}},
  \bibnamefont{and} \bibinfo{author}{\bibfnamefont{N.~J.} \bibnamefont{Cerf}},
  \bibinfo{journal}{arXiv:0710.4858}  (\bibinfo{year}{2007}).

\bibitem[{\citenamefont{Gottesman et~al.}(2001)\citenamefont{Gottesman, Kitaev,
  and Preskill}}]{PhysRevA.64.012310}
\bibinfo{author}{\bibfnamefont{D.}~\bibnamefont{Gottesman}},
  \bibinfo{author}{\bibfnamefont{A.}~\bibnamefont{Kitaev}}, \bibnamefont{and}
  \bibinfo{author}{\bibfnamefont{J.}~\bibnamefont{Preskill}},
  \bibinfo{journal}{Phys. Rev. A} \textbf{\bibinfo{volume}{64}},
  \bibinfo{pages}{012310} (\bibinfo{year}{2001}).

\bibitem[{\citenamefont{Braunstein}(2005)}]{braunstein:055801}
\bibinfo{author}{\bibfnamefont{S.~L.} \bibnamefont{Braunstein}},
  \bibinfo{journal}{Phys. Rev. A} \textbf{\bibinfo{volume}{71}},
  \bibinfo{eid}{055801} (\bibinfo{year}{2005}).

\bibitem[{\citenamefont{Knill et~al.}(2000)\citenamefont{Knill, Laflamme, and
  Viola}}]{PhysRevLett.84.2525}
\bibinfo{author}{\bibfnamefont{E.}~\bibnamefont{Knill}},
  \bibinfo{author}{\bibfnamefont{R.}~\bibnamefont{Laflamme}}, \bibnamefont{and}
  \bibinfo{author}{\bibfnamefont{L.}~\bibnamefont{Viola}},
  \bibinfo{journal}{Phys. Rev. Lett.} \textbf{\bibinfo{volume}{84}},
  \bibinfo{pages}{2525} (\bibinfo{year}{2000}).

\bibitem[{\citenamefont{Kribs and Spekkens}(2006)}]{kribs:042329}
\bibinfo{author}{\bibfnamefont{D.~W.} \bibnamefont{Kribs}} \bibnamefont{and}
  \bibinfo{author}{\bibfnamefont{R.~W.} \bibnamefont{Spekkens}},
  \bibinfo{journal}{Phys. Rev. A} \textbf{\bibinfo{volume}{74}},
  \bibinfo{eid}{042329} (\bibinfo{year}{2006}).

\bibitem[{\citenamefont{Nielsen and Poulin}(2007)}]{nielsen:064304}
\bibinfo{author}{\bibfnamefont{M.~A.} \bibnamefont{Nielsen}} \bibnamefont{and}
  \bibinfo{author}{\bibfnamefont{D.}~\bibnamefont{Poulin}},
  \bibinfo{journal}{Phys. Rev. A} \textbf{\bibinfo{volume}{75}},
  \bibinfo{eid}{064304} (\bibinfo{year}{2007}).

\bibitem[{\citenamefont{Aly and Klappenecker}(2008)}]{isit2008aly}
\bibinfo{author}{\bibfnamefont{S.~A.} \bibnamefont{Aly}} \bibnamefont{and}
  \bibinfo{author}{\bibfnamefont{A.}~\bibnamefont{Klappenecker}}, in
  \emph{\bibinfo{booktitle}{Proceedings of the IEEE International Symposium on
  Information Theory (arXiv:0712.4321)}} (\bibinfo{year}{2008}), pp.
  \bibinfo{pages}{369--373}.

\bibitem[{\citenamefont{Brun et~al.}(2006{\natexlab{a}})\citenamefont{Brun,
  Devetak, and Hsieh}}]{arx2006brun}
\bibinfo{author}{\bibfnamefont{T.~A.} \bibnamefont{Brun}},
  \bibinfo{author}{\bibfnamefont{I.}~\bibnamefont{Devetak}}, \bibnamefont{and}
  \bibinfo{author}{\bibfnamefont{M.-H.} \bibnamefont{Hsieh}},
  \bibinfo{journal}{arXiv:quant-ph/0608027}
  (\bibinfo{year}{2006}{\natexlab{a}}).

\bibitem[{\citenamefont{Brun et~al.}(2006{\natexlab{b}})\citenamefont{Brun,
  Devetak, and Hsieh}}]{science2006brun}
\bibinfo{author}{\bibfnamefont{T.~A.} \bibnamefont{Brun}},
  \bibinfo{author}{\bibfnamefont{I.}~\bibnamefont{Devetak}}, \bibnamefont{and}
  \bibinfo{author}{\bibfnamefont{M.-H.} \bibnamefont{Hsieh}},
  \bibinfo{journal}{Science} \textbf{\bibinfo{volume}{314}},
  \bibinfo{pages}{pp. 436 } (\bibinfo{year}{2006}{\natexlab{b}}).

\bibitem[{\citenamefont{Wilde and Brun}(2008)}]{arx2008wildeOEA}
\bibinfo{author}{\bibfnamefont{M.~M.} \bibnamefont{Wilde}} \bibnamefont{and}
  \bibinfo{author}{\bibfnamefont{T.~A.} \bibnamefont{Brun}},
  \bibinfo{journal}{Phys. Rev. A} \textbf{\bibinfo{volume}{77}},
  \bibinfo{pages}{064302} (\bibinfo{year}{2008}).

\bibitem[{\citenamefont{Wilde et~al.}(2007{\natexlab{d}})\citenamefont{Wilde,
  Krovi, and Brun}}]{arx2007wildeCED}
\bibinfo{author}{\bibfnamefont{M.~M.} \bibnamefont{Wilde}},
  \bibinfo{author}{\bibfnamefont{H.}~\bibnamefont{Krovi}}, \bibnamefont{and}
  \bibinfo{author}{\bibfnamefont{T.~A.} \bibnamefont{Brun}},
  \bibinfo{journal}{arXiv:0708.3699}  (\bibinfo{year}{2007}{\natexlab{d}}).

\bibitem[{\citenamefont{Wilde and Brun}(2007)}]{prep2007wilde}
\bibinfo{author}{\bibfnamefont{M.~M.} \bibnamefont{Wilde}} \bibnamefont{and}
  \bibinfo{author}{\bibfnamefont{T.~A.} \bibnamefont{Brun}},
  \bibinfo{journal}{arXiv:0712.2223}  (\bibinfo{year}{2007}).

\bibitem[{\citenamefont{Barnes}(2004)}]{arx04barnes}
\bibinfo{author}{\bibfnamefont{R.}~\bibnamefont{Barnes}},
  \bibinfo{journal}{arXiv:quant-ph/0405064}  (\bibinfo{year}{2004}).

\end{thebibliography}

\end{document}